# Detection of Deviations in Mobile Applications Network Behavior

L. Chekina, D. Mimran, L. Rokach, Y. Elovici, B. Shapira
Department of Information Systems Engineering and Telekom Innovation Laboratories
Ben-Gurion University of the Negev, Beer-Sheva, Israel

## ABSTRACT
In this paper a novel system for detecting meaningful deviations in a mobile application's network behavior is proposed. The main goal of the proposed system is to protect mobile device users and cellular infrastructure companies from malicious applications. The new system is capable of: (1) identifying malicious attacks or masquerading applications installed on a mobile device, and (2) identifying republishing of popular applications injected with a malicious code. The detection is performed based on the application's network traffic patterns only. For each application two types of models are learned. The first model, *local*, represents the personal traffic pattern for each user using an application and is learned on the device. The second model, *collaborative*, represents traffic patterns of numerous users using an application and is learned on the system server. Machine-learning methods are used for learning and detection purposes. This paper focuses on methods utilized for local (i.e., on mobile device) learning and detection of deviations from the normal application's behavior. These methods were implemented and evaluated on Android devices. The evaluation experiments demonstrate that: (1) various applications have specific network traffic patterns and certain application categories can be distinguishable by their network patterns, (2) different levels of deviations from normal behavior can be detected accurately, and (3) local learning is feasible and has a low performance overhead on mobile devices.

## Categories and Subject Descriptors
C.2.3 [**Computer-Communication Networks**]: Network Operations – *Network monitoring*; H.2.8 [**Database Management**]: Database Applications - *Data Mining*; K.6.5 [**Management of Computing and Information Systems**]: Security and protection – *Invasive software (e.g., viruses, worms, Trojan horses)*.

Security, Algorithms, Performance, Experimentation, Measurement.

## 1. INTRODUCTION
Along with the significant growth in the popularity of smartphones and the number of available mobile applications, the number of malware applications that harm users or compromise their private data has also dramatically increased. Furthermore, the significant growth of social networking and always-connected applications has caused a dramatically increasing influence on traffic and signaling loads on the mobile networks, potentially leading to network congestion incidents.

Network overloads can be caused by either intended attacks or by benign, but unintentionally faulty designed, and thus "network unfriendly" applications. Both the malware activities and the "network unfriendly" applications regularly affect the network behavior patterns and can be detected by monitoring an application's network behavior. Thus, monitoring and analysis of network-active applications' traffic patterns is essential for developing effective solutions for the prevention of network overloads.

The proposed system serves two purposes. First, the system allows protection of mobile device users form malware applications and second, allows for aggregation and analysis of an applications' network traffic patterns (to be used for development of solutions protecting cellular infrastructure from malicious and "network unfriendly" benign applications). Regarding the protection of users form malware applications, the new system supports two main use cases. The first case relates to the applications already installed on a device and the second, to the newly downloaded and installed applications. In the first case, the network traffic pattern of an application can be changed due to: (1) the changes in the user's behavior or (2) an application update to a new benign version or (3) a malicious attack. In this case the system's purpose is to detect the deviation in the application's traffic pattern and correctly classify it to one of the three above mentioned reasons. In the second case, the system's purpose is to identify whether the new application is actually a modification of another application with some new (regularly malicious) behavior.

For the above purposes the new system follows the hybrid Intrusion Detection Systems (IDS) approach and is designed in the client-server architecture. The responsibility of the client-side software is to monitor the applications running on a device, learn their user-specific *local* models and detect any deviations from the observed "normal"[1] behavior. The responsibility of the server-side software is the aggregation of data reported from mobile devices and the learning of *collaborative* models, which represent the common traffic patterns of numerous users for each application. The *local* models are used for detection of deviations in traffic patterns of installed applications; the *collaborative* models are used for verification of newly installed applications vs. the known traffic patterns.

This paper presents the client-side of the whole system, i.e., the sub-system developed for the extraction of per-application traffic features, learning of *local* models, and detection of deviations from the normal user's behavior. This sub-system (will now be referred to as *system*) is implemented as a regular Android application running on the OS user space and evaluated on regular (un-rooted) Android devices. The server-side of the system is currently under development and thus the collaborative model's learning and detection is part of our future research.

---
[1] In this paper we use the term "normal" referring to the regular, non-anomalous observations and do not mean that the data are from the normal distribution in the statistical sense.

In this paper we overview the system components, describe the extracted features, and utilize the machine-learning methods. We then describe the conducted experiments and present the aggregated data analysis and system evaluation results. The performed data analysis reveals that applications have very specific network traffic patterns, and that certain application categories can be distinguishable from their traffic patterns. The system evaluation experiments were conducted with a wide range of different applications, their versions, and several self-developed and real malware applications. The results demonstrate that different levels of deviations from normal behavior can be detected accurately. Specifically, the deviation in up to 20-25 percent of observations might be due to variations in an user's behavior or an application's diverse functionality; deviations in various ranges from 0 up to almost 90 percent of instances might be observed due to an application's version update; lastly, the deviations in 60 and more percent of observations are regularly caused by injected malware. In addition, the conducted experiment proves the feasibility of the proposed implementation and analyses performance overhead on mobile devices.

The rest of the paper is organized as follows. Section 2 discusses related work. Section 3 describes the client-side system components and methods. Section 4 presents the results of the evaluation experiment. Next, in Section 5 we discuss the achieved results. Lastly, Section 6 concludes the paper and outlines future research.

## 2. RELATED WORK

Traditionally Intrusion Detection Systems are classified according to the protected system type as either host-based (HIDS) or network-based (NIDS) [1]. A network-based IDS is located on a central or distributed dedicated server(s) and monitors any number of hosts. Its performance is based on analysis of network related events, such as traffic volume, IP addresses, service ports, protocol usage, etc. Traffic monitoring is usually accomplished at concentrating network units, such as switches, routers, and gateways. On the other hand, a host-based IDS resides on and monitors a single host machine. Its performance is based mainly on an analysis of events related to OS information, such as file system, process identifiers, system calls, etc. [8].

The current research work is unique in the IDS field in the sense that it proposes a host-based system whose performance is based on application-level network events solely. The anomalous behavior of an application is detected in real time on the device based on the observed network traffic patterns. This approach is justified by the fact that many malware applications use network communication for their needs, such as sending a malicious payload or a command to a compromised device, or getting user's data from the device. In fact, a recent survey of mobile malware reveals that about 70% of known malware steals user's information or credentials [7]. Such types of behavior influence the regular network traffic patterns of the application and can be identified by learning the application's "normal" patterns and further monitoring network events.

Recently, with the dramatic increase in the number of malware applications targeting smartphones, various methods for intrusion detection on mobile devices have been proposed. A review of several such methods are presented in [2,18]. Most of the IDSs for mobile devices have focused on host-based intrusion detection systems applying either anomaly- or rule-based methods on the set of features that indicate the state of the device [17]. However, in most cases, the data interpretation processes are performed on remote servers motivated by limited computational resources of the mobile phone. Only a few of the proposed systems perform the learning or data analysis directly on the device [6, 10, 19] and even less have applied statistical or machine-learning techniques [10,19], even though such techniques are very popular and have been successfully used in traditional anomaly detection systems [8, 19]. Most of the systems either send the observed data to the server for analysis [2, 4, 12, 14, 16, 22] or perform the learning process offline on the server and plant the learned models back to the devices for the detection process [15, 17, 18]. Differently, from the earlier proposed methods, our system performs application-based anomaly detection using only application-level network traffic features while both learning and detection processes utilize the machine-learning algorithms and are performed on the device.

Consider the earlier proposed systems where learning is performed on the mobile devices (as this is the most close to our work). The system proposed by Shamili et al. in [19] utilizes a distributed Support Vector Machine algorithm for malware detection on a network of mobile devices. The phone calls, SMSs, and data communication related features are used for detection. During the training phase support vectors (SV) are learned locally on each device and then sent to the server where SVs from all the client devices are aggregated. Lastly, the server distributes the whole set of SVs to all the clients and each of the clients updates his own SVs. Thus, although a part of the learning is performed on the device, the server and communication infrastructure, along with additional bandwidth load, are required. Our approach, though planned as a part of the wider client-server system, can perform and be utilized as a stand-alone solution running independently on each mobile device. Once more, authors evaluated their methods using the MIT Reality dataset [5] with manually injected symptoms of malware behavior and no estimation of resource overhead was carried out.

Li et al. [10] presented an approach for behavior-based multi-level profiling IDS considering telephony calls, device usage, and Bluetooth scans. They proposed a host-based system which collects and monitors user behavior features on a mobile device. A Radial Basis Network technique was used for learning profiles and detecting intrusions. However, the system capabilities were, also, tested offline only using the MIT Reality dataset and its feasibility on mobile devices was not tested or verified.

Therefore, our work is one of the first practical implementations of the machine-learning induction algorithm for mobile OS in general and for Android platform specifically.

## 3. SYSTEM OVERVIEW

This section describes the main functional units of our system responsible for the online detection of traffic deviations on mobile devices. The system's architecture is presented in Figure 1 and consists of the following main components:

*Graphical User Interface* (GUI) – is responsible for communication with user; presents the relevant information, receives the desirable parameters configuration, starts and stops the monitoring, etc.

*Alerts Handler* – is responsible for presenting the alerts to the user interface and processing the user's response.

*Features Extraction* – performs the measurements of the defined features at the specified time periods.

*Features Aggregation* – computes the defined aggregations over all the extracted measurements for the specified time period.

*Local Learner* – induces the local models representing an application's traffic patterns specific for the user.

*Anomaly Detector* – is responsible for the online analysis of an application's network behavior and detection of deviation from its normal patterns.

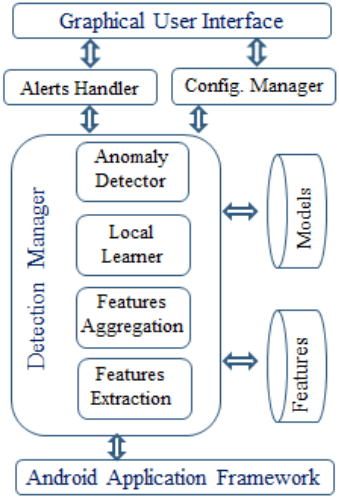

**Figure 1. System architecture.**

In the following paragraphs we describe the four main logical components; Features Extraction, Features Aggregation, Local Learner, and Anomaly Detector modules.

### 3.1 Features Extraction

As mentioned above, the Features Extraction module is responsible for extraction (i.e., measuring and capturing) of the defined list of features for each running application at each defined time period. For this purpose it uses the APIs provided by the Android Software Development Kit (SDK). Below is a list of the currently extracted features:

- sent\received data in bytes and percent;
- network state (Cellular, WiFi or "No network");
- time (in seconds) since application's last send\receive data;
- send\receive mode (eventual\continuous) – derived from "since-last-send\receive-seconds", i.e., if the last send or receive data event was detected less than a specified number of second ago, the corresponding (send or receive) mode is continuous, otherwise it is eventual;
- two application states – the first, specifies whether the application is in foreground or background and the second, specifies whether the application is among the active or non-active tasks at the time of the measurement;
- time in fore\background (in seconds and percent) - total time that an application has been in fore\ background since the last monitoring of this application was started;
- minutes since application's last active\modified time.

Additionally, the following features are planned to be extracted in the next version of the system:

- number of total\concurrent connections of the application;
- number of sent\received TCP\UDP packets;
- number of sent\received TCP\UDP payload bytes;
- number of sent\received TCP segments.

The extraction time period is a configurable parameter. For the initial experiments with the system we set it to 5 seconds, however it is subject to change according to the results of future evaluation experiments.

### 3.2 Features Aggregation

The purpose of the Features Aggregation module is to provide a concise representation of the extracted application's traffic data. For this purpose, a list of various aggregation functions was defined. The instances of the aggregated data are used to induce machine-learning models representing an application's behavior and for further anomalies detection. To get a notion of the usefulness of the various features for our problem, in the current work an extended list of possible aggregated features was defined and evaluated. According to the evaluation results, a preliminary list of the few most useful features is determined in the Evaluation Section. Below is a list of all the currently defined and aggregated features:

- Average, standard deviation, minimum, and maximum of sent\received data in bytes;
- Average, standard deviation, minimum, and maximum of sent\received data in percent;
- Percent of sent\received bytes;
- Time intervals between send\receive events - the send\receive events that occurred within the time interval of less than 30 seconds from the previous corresponding event contribute to the calculation of the *inner average send\receive time interval*. The events that occurred within the time interval above or equal to 30 seconds from the previous corresponding event contribute to the calculation of the *outer average send\receive time interval*. Additionally, two types of intervals; *local* – for each specific aggregation time period, and *global* – averaged over the whole monitoring process, were calculated. The *local* time intervals describe an application's behavior at certain monitoring time points, while the *global* time intervals describe the application's general behavior observed up until the current point of time;
- Network state - Cellular, WiFi, none or mixed. The mixed state was determined in the case where several different states (i.e., Cellular and WiFi) were observed during the same aggregation period;
- Minutes past since application's last send\receive data event;
- Application state 1- foreground, background or mixed. Mixed state was determined in the case where several different states were observed during the same aggregation period;
- Application state 2 - active, non-active or mixed;
- Total and local time (in seconds) for which the application was in the fore\background state. Local time may vary from 0 to 60 seconds and represent the value specific for the current aggregation interval, while the total time is aggregated over the whole application's active time period;
- Minutes past since the application's last active time;
- Days past since application's last modified time determined according to the application's installer file (i.e., ".apk" for Android) modification time.

Similar to the extraction, the aggregation time period is a configurable parameter. For the current experiment it was set to 1 minute.

### 3.3 Local Models Learning

Our main goal in this paper is to learn user specific network traffic patterns for each application and determine if meaningful changes occur in the application's network behavior. This task relates to the family of semi-supervised anomaly detection problems, which assumes that the training data has samples for "normal" data examples only. These types of problems can be solved, for example, with one-class support vector machines (SVMs), the local outlier factor (LOF) method, or clustering based techniques [3, 13].

In this paper we decided to convert the semi-supervised learning problem into a set of supervised problems for which numerous well established and quick algorithms exist. For this purpose we follow the "cross-feature analysis" approach presented in Huang et al. in [9] and then further analyzed by Noto et al. in [13]. Both of these works have found this approach successful and useful for anomalies detection. However, [9], only considers features with discrete values and [13] mainly focuses on methods for combining the results of multiple feature predictors to make the final decision about instance normality. In our problem most of the features are numerical and an efficient implementation is desired as it is supposed to run on mobile phones. Thus, in this work we follow the general idea of the above approach; however, our implementation differs in several details. Both the general idea and our current implementation are presented below.

The main assumption underling the "cross-feature analysis" approach is that in normal behavior patterns, strong correlations between features exist and can be used to detect deviations caused by abnormal activities. The basic idea of a cross-feature analysis method is to explore the correlation between one feature and all the other features. Formally, it tries to solve the classification problems $C_i: \{f_1, \ldots, f_{i-1}, f_{i+1}, \ldots, f_L\} \to \{f_i\}$, where $\{f_1, f_2, \ldots, f_L\}$ is the features vector and $L$ is the total number of features. Such a classifier is learned for each feature $i$, where $i = 1, \ldots L$. Thus, an ensemble of learners for each one of the features represents the model through which each features vector will be tested for "normality". The C4.5 Decision Tree algorithm was used for learning the classification model when the target class consists of categorical values. For learning the classification model for the numeric target attribute, several methods capable with numeric classes were evaluated. The evaluated methods and their results are presented in the Evaluation Section.

### 3.4 Anomaly Detection

This module is responsible for the online analysis of an application's network behavior and the detection of deviations from normal patterns. The procedure utilized for testing each individual instance is further described.

When a feature's vector representing a normal event is tested against $C_i$, there is a higher probability for the predicted value to match (for discrete features) or be very similar (for numeric features) to the observed value. However, in the case of a vector representing abnormal behavior, the probability of such a match or similarity is much lower. Thus, by applying all the features models to a tested vector and combining their result, a decision about vector normality can be derived. The more different the predictions are from the true values of the corresponding features, the more likely that the observed vector comes from a different distribution than the training set (i.e., represents an anomaly event).

For each predictor $C_i$ we compute the probability of the corresponding feature value of a vector $x$ to come from a normal event. This probability, noted $P(f_i(x) \text{ is normal})$, is calculated as $1 - distance(C_i(x), f_i(x))$, where $C_i(x)$ is the predicted value and $f_i(x)$ is the actual observed value. The *distance* between two values for a single numeric feature is the difference in actual and predicted values divided by the mean of the observed values for that feature. If the difference is higher than mean value, the distance is assigned with a constant large value (such as 0.999). The distance for a discrete feature is the Hamming distance (i.e., 1 if the feature values are different and 0 if they are identical). To get the total probability of a vector x to represent a normal event, we make a naïve assumption about the sub-model's independence and multiply[2] all the individual probabilities computed for each one of the feature values. A threshold distinguishing between normal and anomalous vectors is learned during algorithm calibration on the data sets with labeled samples.

However, detection of abnormality in a single observed instance is not sufficient to determine whether that application's behavior has been meaningfully changed. Such sole anomalies can be caused by changes or noise in a user's behavior. In order to reduce the False Alarms rate and improve the effectiveness of the proposed system in general, we want to define a procedure which considers the consequent observations and derives a decision comprised of the individual predictions for each one of these observations. For example, an alarm can be dispatched only when an anomaly was detected in a certain number of consequent instances (i.e., 3 consequent instances were detected as anomalous) or when an anomaly was detected in a certain percent of instances during a specified time period (i.e., 3 or more anomalies during a 10 minute interval). The exact procedure has been determined according to our observations during the evaluation experiments and is described in the next section.

### 4. EVALUATION

This section presents the initial analysis of applications' traffic patterns and an evaluation of the proposed detection system. First, the research questions that we attempt to answer are described in Section 4.1. Section 4.2 describes the data collected for the experiments. In Section 4.3 we present the traffic patterns observed for several popular applications. In Section 4.4 the system calibration and evaluation processes are described, and the results observed for three types of the tested software; regular applications, self-written malware, and real malware, are presented. Lastly, in Section 4.5, the overhead of the proposed system on mobile phone resources is estimated.

---

[2] In this paper we utilize this method due to its simplicity and computational efficiency, despite the known incorrectness of the underlying independence assumption. Utilization and analysis of more sophisticated methods is one of our future tasks.

## 4.1 Research questions

The performed evaluation experiments aimed to answer the following research questions:

1. Is it possible to model an application's network behavior so that any deviation from its normal behavior would be detected?
2. Which network application-level features are most useful for modeling an application's network traffic patterns?
3. Which classification algorithm is the most effective for learning the models and detection of anomalous behavior when most of the features are numeric values?
4. What level of detection accuracy (and false alarms) could be reached using application level network-behavioral features only?
5. How much overhead on mobile phone resources is caused by applying machine-learning and detection algorithms directly on the device?

## 4.2 Data collection

For the initial data analysis and evaluation of the proposed method, the Features Extraction module was installed and ran on the personal Android devices of eight volunteer users. During this period the applications' features were extracted and aggregated, as described in Sections 3.1 and 3.2. We have from 2 weeks up to 3 months of data for each user.

Additionally, we experimented with one self-written and five real Trojan malware applications. Each evaluated malware application has two versions: the original application asking for network access permission for benign purposes (such as displaying advertisements, best scores updated, etc.) and the repackaged version of the original application with injected malware code performing network connections for malicious purposes.

For the self-written malware tests, the "Snake" game scenario from [17] was utilized. The benign application is actually an offline game financed through mobile advertisements and uses the network for the advertisements only. The infected version of this application is added with a background service which quietly takes pictures and sends them off to a remote server while the user is playing a game.

For the tests with the real malware, five infected applications and their benign versions were utilized. The infected applications and the corresponding versions of the benign application were retrieved from a repository collected by crawling the official and various alternative Android markets for over a year and a half. We used two applications injected with PJApps [20] Trojan; Fling and CrazyFish, two applications injected with Geinimi [21] Trojan; Squibble Lite and ShotGun, and one sample of DroidKungFu-B [11] malware found within the OpenSudoku game.

The PJApps Trojan, which was discovered in applications from unofficial Android marketplaces, creates a service that runs in the background, sends sensitive information containing the IMEI, Device ID, Line Number, Subscriber ID, and SIM serial number to a web server, and retrieves commands from a remote command and control server.

The Geinimi Trojan arrives on the device as part of repackaged version of legitimate applications. The applications repackaged with Geinimi Trojan have been found in a variety of locations, including unofficial marketplaces, file-share sites, and miscellaneous websites. When installed, the Trojan attempts to establish contact with a command and control server for instructions and once the contact is established, it transmits information from the device to the server and may be instructed to perform certain actions.

The DroidKungFu-B is a version of the DroidKungFu malware. The initial DroidKungFu malware is known for its capability of rooting Android phones with OS 2.2 or below. The infected applications were found among alternative Android markets targeting the Chinese audience. The DroidKungFu-B version targets already rooted phones and requests for the root privilege. In either case (with or without the root privilege), the malware collects and steals the phone information (e.g., IMEI, phone model, etc.).

The malware applications and their benign counterparts were executed on a specially designated device and their behavior was collected and analyzed.

## 4.3 Data analysis

This section presents the traffic patterns observed while analyzing the collected data of several popular applications with heavy network usage, such as Facebook, Skype, Gmail, and WhatsApp. Although the graphs are presented in two dimensions only, average sent vs. average received bytes, the distinguishable patterns of each application are clearly highlighted. The graphs representing the network behavior of the above applications on the devices of different users are presented in Figures 2(a-d) correspondingly. The data points of different users are plotted in different colors.

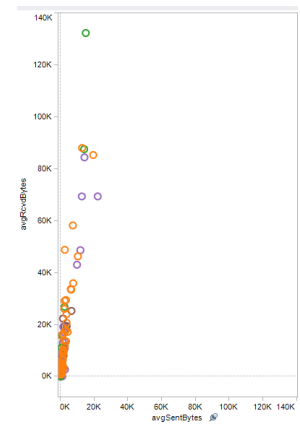

**Figure 2a. Facebook.**

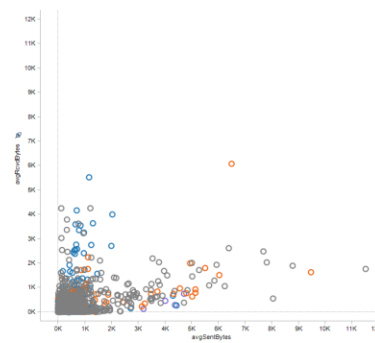

**Figure 2b. Skype.**

As can be seen from the graphs, each one of the analyzed applications has its own specific traffic pattern which is easy distinguishable from other applications. Note that on each of the graphs, the axis value's range is different. Additionally, other features can be utilized for differentiation in less certain cases.

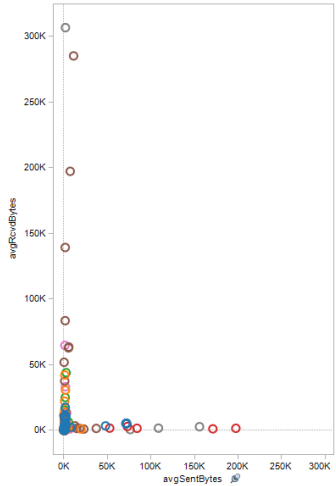

**Figure 2c. Gmail.**

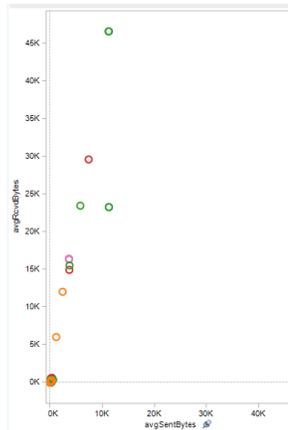

**Figure 2d. WhatsApp.**

Next, the two graphs presented in Figures 3a and 3b depict the behavior of different applications from the same type. Figure 3a depicts the traffic pattern of two e-mail client applications: Gmail and Android's native Email client. Figure 3b depicts the traffic pattern of two Internet browsers: Mozilla Firefox and device's native Browser application. Data points of different applications are plotted in different colors.

It can be seen from the graphs that different applications from the same functionality type have very similar traffic patterns among them, while the traffic patterns of various application types are different.

All the above observations lead us to the following conclusions:

1. Modeling a mobile application's network behavior using application-level features only is possible;
2. The proposed approach is feasible: applications have certain patterns of their normal behavior, which can be learned so that any meaningful deviations from these patterns would be detected;
3. Additionally, the observed network behavior of an application can be used to determine whether this application is what it claims to be, given that normal patterns of this application are known;
4. Certain types of applications have similar network traffic patterns which can be used, for example, for traffic classification or hierarchical clustering of applications.

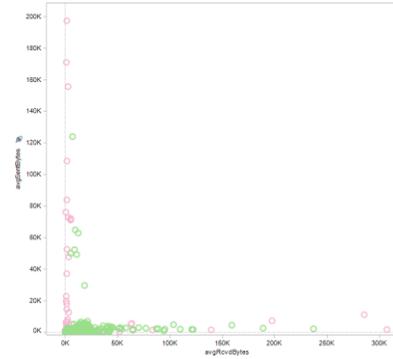

**Figure 3a. E-mail clients.**

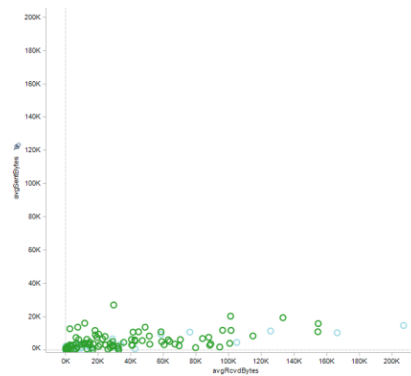

**Figure 3b. Internet browsers.**

### 4.4 System evaluation

We conducted two types of experiments. The first type, which can be considered as a calibration experiment, served several purposes: 1) selection of optimal features subset, 2) evaluation of several machine-learning algorithms as our base learners, 3) determination of the minimal sufficient training set size, and 4) determination of the strategy for raising the "Anomaly" alarm in case one or more anomalous records are detected.

For evaluation of different classification algorithms and features selection, the following standard measures were employed: True Positive Rate (TPR) measure (also known as Detection Rate), which determines the proportion of correctly detected changes from an application's normal behavior; False Positive Rate (FPR) measure (also known as False Alarm Rate), which determines the proportion of mistakenly detected changes in an actually normal application behavior; and Total Accuracy, which measures the proportion of a correctly classified application behavior as either anomalous or normal.

The purpose of the second experiment type, which can be considered as a test experiment, was to evaluate the ability of the proposed system to distinguish between benign and malicious versions of the same application and between two benign yet different versions of the same application. Additionally, the low False Alarm rate on the data records of the same application version was verified.

In this section we first describe the calibration experiments and their results followed by the test experiments and their results.

### 4.4.1 Calibration experiments

For the calibration experiments a set of 16 datasets were extracted and prepared from the collected data. Each one of the 16 datasets consists of train and test records. The datasets were selected and prepared in the following way. In half of the datasets (i.e., in 8 datasets) both the train and test records were taken from the same version of a certain application. These datasets were used to verify a low detection rate on the records of the same application and determine the deviation level in traffic patterns that can be attributed to the application diversity and changes in a user's behavior. In the other 8 datasets, train and test records were taken from different versions of a certain application. These datasets were used to verify the higher detection rate than seen in the cases with the same application version. However, in some cases, the low detection rate for the different application versions is acceptable, as different application versions are not obligated to contain any network related updates. For both, the calibration and test experiments, the train size for all applications was limited to the maximum of 150 instances, and the test size to the maximum of 400 instances. On datasets with fewer available examples, the full train and test sets were utilized.

#### 4.4.1.1 Features selection

A wide range of features have been defined and presented in Section 3. However, extraction and aggregation of a large number of features on a mobile device is a very inefficient and resource wasting process. Additionally, learning classification models and detection with a large number of features is much more computationally expensive. Furthermore, the presence of redundant or irrelevant features may decrease the accuracy of the learning algorithm. Thus, our purpose in the features selection is to identify a minimal set of the most useful features. There are several groups of features among the defined list of aggregated features for which extraction and calculation is performed together using the same amount of resources. Thus, reducing one or a few features from such a group, while at least one feature from such a group has to be calculated, will not reduce the extraction and calculation overhead significantly. The standard approaches for features selection, such as Filter and Wrapper, are not applicable in this case, as they cannot consider the above described constraints between the features. For this reason, twenty feature subsets of various sizes and containing various groups of features were manually defined. The threshold distinguishing between the normal and anomalous vectors was defined separately for each one of the features subset in the preliminary calibration experiments, as it depends on the number and type of the involved features.

#### 4.4.1.2 Evaluated base learners

Considering the prevalence of numerical attributes among the defined aggregated features, and the resource consumption issue, we decided to evaluate the following classifiers as candidates for our base-learner algorithm: Linear Regression, Decision Table, Support Vector Machine for Regression, Gaussian Processes for Regression, Isotonic Regression, and Decision/Regression tree (REPTree). The Weka [23] open source library was used for evaluation of these algorithms.

All the defined feature subsets were tested with all the evaluated base learning algorithms on the calibration datasets described above.

As was previously mentioned, sometimes abnormal instances can be caused by either changes in a user's behavior or by diversity in an application's functionality. In order to determine the acceptable rate of such abnormal instances in a normal application's behavior, we evaluated the possible range between 5 and 25 percent with step 5. Thus, the results of all the tested algorithms and feature subsets were evaluated for 5 different "anomaly acceptance" rates; 5, 10, 15, 20, and 25.

The results of the calibration experiments reveal the two best combinations of the base learning algorithm and features subset. The two best base algorithms are the Decision Table and the REPTree. The two best features subsets, presented in Table 1, are very similar to each other; one of the subsets includes all the features from another plus two additional features.

**Table 1. Selected features subsets**

| Features Subset #1 | Features Subset #2 |
|---|---|
| Avg. Sent Bytes | Avg. Sent Bytes |
| Avg. Rcvd. Bytes | Avg. Rcvd. Bytes |
| Pct. Of Avg. Rcvd. Bytes | Pct. Of Avg. Rcvd. Bytes |
| Inner Avg. Send Interval | Inner Avg. Send Interval |
| Inner Avg. Rcvd. Interval | Inner Avg. Rcvd. Interval |
| Outer Avg. Send Interval | Outer Avg. Send Interval |
| Outer Avg. Rcvd. Interval | Outer Avg. Rcvd. Interval |
| - | Avg. Sent data Percent |
| - | Avg. Rcvd. data Percent |

As can be seen from the Table above, there are seven features included in both of the best subsets. Thus, we can conclude that these features are the most useful for modeling application's network traffic.

As for the estimated algorithm accuracy performance, the Decision Table algorithm in conjunction with the features subset #1 and "anomaly acceptance" rate 20 percent results in TPR=0.8, FPR=0, and Total Accuracy=0.875 and the REPTree algorithm in conjunction with the features subset #2 and "anomaly acceptance" rate 25 percent demonstrates exactly the same accuracy values.

For a better insight into the detection rate observed in the calibration datasets, the results of the Decision Table algorithm in conjunction with the features subset #1 and the REPTree algorithm in conjunction with the features subset #2 are presented in Table 2 (errors are marked in bold font).

**Table 2. Detection rate on calibration datasets**

| Application Name | Detected anomalous records (%) | |
|---|---|---|
| | Decision Table | REPTree |
| *Different application versions* | | |
| twitter | 60.9 | 91.3 |
| groupme | 74.5 | 82.9 |
| gmail | **5.0** | **11.9** |
| facebook | 25.8 | **17.3** |
| twitter | **1.6** | 26.0 |
| firefox | 26.8 | 32.8 |
| whatsapp | 29.2 | 44.4 |
| linkedin | 32.0 | 48.0 |
| *Same application version* | | |
| twitter | 0.0 | 6.7 |
| facebook | 1.3 | 3.9 |
| groupme | 6.7 | 10.0 |
| gmail | 16.0 | 8.0 |
| twitter | 1.0 | 14.0 |
| firefox | 20.0 | 20.0 |
| whatsapp | 7.5 | 13.5 |
| whatsapp | 10.5 | 6.5 |

It can be seen that for most of the different application versions the detection rate is above the determined "anomaly acceptance" rate of 20-25 percent for both algorithms. At the same time, the detection rate on the test sets from the same application version is always below 20 percent. Thus, the detection strategy consisting of several steps can be defined as follows: 1) raise the "Anomaly Alarm" if at least 3 consequent abnormal instances are detected, 2) raise the "Anomaly Alarm" if at least 3 abnormal instances are detected among the five consecutive observations, 3) raise the "Anomaly Alarm" if at least 3 abnormal instances are detected among the ten consecutive observations. According to this strategy, a system will raise an alert about any meaningful changes in an application's network patterns, including those caused by a version update. Further on the version update can be verified within the mobile OS and the Alert with the relevant information (including abnormal instances rate, whether a version update was detected and when) can be presented to the user.

### 4.4.1.3 Training set sizes analysis

An important question regarding the proposed detection system is how quickly the detection can be started (i.e., how many examples are needed for sufficient learning of the network traffic patterns)? To answer this question we evaluated the two winning algorithms using train sets of various sizes. This experiment was executed on all the calibration datasets, varying the train set size from 10 to 100 or the maximum of the available instances with step 10, and from 100 to 400 with step 25.

The results with both algorithms show that, in most cases, the train size of 30-50 examples is sufficient for learning a stable model which is able to determine the level of deviation between an application's traffic patterns correctly. However, it was found that in several cases, for such diverse applications like Facebook and Gmail, a larger amount, such as 80 – 150 examples, is needed for learning a stable model. Considering the fact that in the current experiments each data instance represents one minute of an application's network usage, we conclude that a relatively short time, varying from 30 minutes to 2.5 hours of network activity is required for our system to learn the patterns of a new application. Note that certain applications with rare network usage may actually require much longer time, while the required amount of network behavior data is aggregated.

Additionally, we hypothesize that the minimal size of the training set required for learning stable models can be automatically predicted based on the meta-features of the observed data (i.e., range of values, variance, etc.) Investigation in this direction is a part of our future work.

### 4.4.2 Evaluation results

To test the proposed system, a set of other 12 datasets, 6 with train and test records from the same application version and 6 with train and test records from different application versions, was used. Additionally, the system was tested with one self-written and five real malware applications, as described in Section 4.2.

The detection rate of the Decision Table and REPTree algorithms in conjunction with the features subset #1 and #2 correspondingly, on the evaluated datasets are presented in Table 3 (detection errors are marked in bold).

It can be seen that for all the malware applications, the high level deviations (60-100%) were detected. Furthermore, deviations at various levels were detected in most cases when the learned models were tested with instances from a different application version. The undetected versions of Facebook and WhatsApp applications can be explained by very few or no network-related changes in the considered application versions. Additionally, the detection rate for all the cases when the learned models were tested with instances from the same application version are below the defined "anomaly acceptance" rate of 20 percent for the Decision Table algorithm and of 25 percent for the REPTree algorithm. Thus, using the Decision Table algorithm's "anomaly acceptance" rate of 20 percent, the estimated method's accuracy on the test data is the following: TPR=0.82, FPR=0.0 and Total Accuracy=0.875. For the REPTree algorithm with the determined "anomaly acceptance" rate of 25 percent, the estimated accuracy on the test data is even higher: TPR=0.91, FPR=0.0, and Total Accuracy=0.94.

**Table 3. Detection Rate on test datasets**

| | Application Name | Detected anomalous records (%) | |
|---|---|---|---|
| | | Decision Table | REPTree |
| Regular applications | *Different versions* | | |
| | twitter | 57.8 | 62.2 |
| | twitter | 78.2 | 34.8 |
| | **facebook** | **0.5** | **3.3** |
| | groupme | 80.9 | 87.2 |
| | whatsapp | **16.7** | 28.9 |
| | *Same version* | | |
| | groupme | 0.0 | 0.0 |
| | groupme | 0.0 | 15.0 |
| | gmail | 14.8 | 22.2 |
| | facebook | 16.0 | 15.7 |
| | firefox | 20.0 | 22.8 |
| Malware applications | *Self-written malware* | | |
| | Snake | 100.0 | 100.0 |
| | *Real malware* | | |
| | Fling | 63.6 | 66.8 |
| | OpenSudoku | 100.0 | 100.0 |
| | ShotGun | 97.0 | 89.5 |
| | Squibble | 90.0 | 95.0 |
| | Crazy Fish | 100.0 | 100.0 |

Consider the surprisingly high detection rate in the several real malware applications. Note that in the self-written malware the 100% detection rate is not surprising, as the benign and malicious versions are significantly different in their network usage patterns. However, in the case of the real malware applications, the 100% detection rate is not obvious. In the applications infected with the Trojans, the main application's functionality is preserved and some new functionality is added. Thus, some part of the data related to the old functionality might be expected to remain unchanged. This is actually the case with the Fling application where online mobile advertisements are displayed while the application is in the phone's frontend in both versions. Thus, the records corresponding to the time when the game was actually played were less affected by the Trojan functionality and thus the observed detection rate is "only" 60%. Analysis of the data aggregated from the benign and malicious versions of the evaluated applications shows that the significant differences are caused by a background process that is running even when an application is not active and performs multiple connections (or connection attempts) with the server at constant time intervals. This behavior has a significant effect on such features, such as avg. sent\received bytes, number of sent\receive events, global outer\inner sent\receive intervals, and others. Most of the mentioned and significantly influenced features are contained in the utilized features subsets and this explains the high detection rate.

## 4.5 Resources Overhead

This section evaluates the overhead caused by the learning and detection processes on mobile phone resources in terms of memory consumption, CPU load, and time needed for a model's induction and vector testing processes. Experiments were performed on a Samsung Galaxy S GT-i9000 running Android OS version 2.2. One of the selected combinations, the REPTree algorithm in conjunction with the features subset #2, was used for the overhead evaluation experiments.

Note that online monitoring is performed for network-active applications only. Generally there are no more than 2-3 such applications running simultaneously on a device most of the time. Additionally, we assume that during the time periods of a user's normal activity, the number of such applications may reach no more than 10 – 15 network-active concurrent processes. Thus, for performance estimation, we consider a scenario of 10 concurrently monitored applications. We estimate the memory and CPU load for learning the 10 application models and further constant monitoring of their network traffic. For a better estimation of memory consumption, the results were averaged through 10 distinct experiments.

### 4.5.1 Memory consumption

The memory consumption of the application changes in intervals from 7,035 KB±8 before the learning process has started to 7,272 KB±15 after storing the 10 learned models in memory (which is approximately 1.4% of the device's RAM). Storage of each additional model in memory consumes about 24 KB±0.7 on average. For comparison the memory consumption observed for several constantly running Android services and other popular applications is presented below: Android System – 24,375 KB; Phone Dialer – 8,307 KB; Antivirus – 7,155 KB; TwLauncher – 22,279KB; Email – 10,611 KB; and Gmail – 9,427 KB. The detection process has no effect on the consumed memory.

### 4.5.2 CPU consumption

The CPU consumption peaks occurred at the times of the actual model's learning and were in the interval of 13% ±1.5. Note that model's learning operations occur very rarely, either when a new application is installed or when a model's update is needed (due to a new application version or changes in user's behavior). The CPU consumption observed during the process's idle time was in the interval of 0.7% ±1.02. Time needed to learn a model (using 50 training examples) varies in intervals of 249 msec. ± 27.4.

The time needed for testing a single instance varies in intervals 3.6 msec. ± 2.5. Recall that aggregated features vectors are tested once at the defined aggregation time interval (one minute for these experiments). The CPU consumed by testing 10 concurrent instances (one for each one of the assumed active applications) varies in intervals of 1.8% ± 0.8.

Note that the results of this experiment depict the resources' overhead caused during the user's high activity time periods. During the, presumably much longer time periods of the user's normal activity, an even lower overhead is expected.

## 4.6 Discussion

In this section the defined research questions are discussed in light of the experimental results.

Considering the first question, it has been shown that modeling a mobile application's network behavior using application-level features only is possible and that the proposed approach is feasible; patterns of the normal application's behavior for an application can be learned and then any meaningful deviations from these patterns can be accurately detected.

In regards to the second question, 7 network application-level features were found as the most useful for modeling an application's network traffic patterns. Some further fine tuning evaluations to add\remove certain features could probably slightly improve the achieved results.

As for the most effective classification algorithm to be used as a base-learner in the "cross-feature analysis" approach, the REPTree and Decision Table algorithms were found to be the most successful. The high detection accuracy utilizing these algorithms was confirmed on the test data as well.

Considering the forth question, it was shown that a high True Positive Rate along with zero False Alarms could be achieved using the selected classification algorithms and features subset. An interesting question for future research is whether this accuracy could be improved using the network-level features in addition to the currently selected features?

We have also demonstrated that the proposed online learning and detection has relatively low overhead on the mobile phone resources, comparatively to the resources consumed by other permanently running services, and thus is acceptable for running on smartphones. Unfortunately, we could not evaluate the Features Extraction and the Aggregation processes' impact on the mobile phone resources due to the fact that an extended list of features was observed and calculated. In our future work we plan to re-implement the Features Extraction components retaining only the most effective features and thus the performance overhead of the whole system can be determined.

## 5. SUMMARY AND CONCLUSIONS

In this paper we presented a novel system for detecting meaningful deviations in a mobile application's network traffic patterns. The presented system, although initially planned as a host-based part of a larger hybrid (i.e., client-server) system, is a fully-functioning stand-alone monitoring application for mobile devices, which can be used alone or in conjunction with other methods. One of the main capabilities of the proposed system is protection of mobile device users form malicious attacks on their phones. The detection is performed based on the application's network traffic patterns only.

Machine-learning methods were used for learning and detection purposes. This work represents one of the first attempts to run learning and detection processes on the device itself which demonstrates the feasibility and acceptable resources overhead of the proposed method. Although the overhead of the Features Extraction process was not yet measured, this is an unavoidable part of a host-based system, and thus those processes should be of maximal efficiency in any case. An estimation of the overhead of the whole system is one of our upcoming tasks.

Experimentally a subset of few network application-level features most useful for modeling traffic patterns were identified among the wide range of the extracted and aggregated features. Additionally, several classification algorithms suitable for handling numerical data were evaluated and the two most effective methods were selected. Moreover, the traffic patterns of several popular applications were presented and analyzed. It was shown that many applications have very specific network traffic patterns and that certain application's categories can be distinguishable from their network traffic patterns.

Results of the evaluation experiments conducted with different benign and malware applications demonstrate that a high True Positive Rate along with low False Alarms could be achieved using the proposed method. Specifically, it was shown that different levels of deviations from normal behavior can be detected accurately. Thus, the deviations in up to 20-25 percent were observed in tests with the same application versions. These deviations can be explained by different user's behavior and the diverse functionality provided by certain applications. Deviations in various ranges from 0 up to almost 90 percent of instances might be observed in the different application's version. Such a wide deviations range is explained by different levels of network related changes in the new versions. The deviations in 60 and more percent of records were observed in the applications containing an injected malware. Summarizing the results of the presented research work, we conclude the proposed method is feasible and effective for the detection of different deviation levels in the application network patterns.

Some of our main future research directions are: verifications on whether the detection accuracy can be improved using other application- and network-level features; automatic prediction of the minimal training set size sufficient for learning stable models, based on the meta-features of the observed data; development and analysis of methods for a collaborative model's learning and detection.